\documentclass[prl,twocolumn,supplplerscriptaddress]{revtex4-2}

\usepackage{amsfonts}
\usepackage{amssymb}
\usepackage{amsmath}
\usepackage{graphicx}
\usepackage{dcolumn}
\usepackage{bm}
\usepackage{units}
\usepackage{multirow}
\usepackage{CJKutf8}
\usepackage{subfigure}
\usepackage{color}%
\usepackage{url}
\usepackage{soul}
\usepackage[colorlinks,linkcolor=blue,anchorcolor=blue,citecolor=blue,urlcolor=blue]{hyperref}

\begin{document}


\title{RKKY-like interactions between two magnetic skyrmions}

\author{X. C. Hu$^{1}$}
\author{H. Y. Yuan$^{2}$}
\author{X. R. Wang$^{1}$}
\email[Corresponding author: ]{phxwan@cuhk.edu.cn}
\affiliation{$^{1}$School of Science and Engineering, Chinese University of Hong Kong (Shenzhen), Shenzhen, 51817, China}
\affiliation{$^{2}$Institute for Advanced Study in Physics, Zhejiang University, 310027 Hangzhou, China}
\date{\today}

\begin{abstract}
Understanding skyrmion-skyrmion interactions is crucial for effectively manipulating the motion of multiple skyrmions in racetrack and logic devices. However, the fundamental nature and microscopic origins of these interactions remain poorly understood. In this study, we investigate skyrmion-skyrmion interactions in chiral magnetic films and reveal that they possess intrinsic, anisotropic, and oscillatory characteristics. Specifically, we demonstrate that the attractive and repulsive forces between skyrmions oscillate with a well-defined period, akin to the Ruderman-Kittel-Kasuya-Yosida (RKKY) coupling observed between two magnetic moments in metals. Our analysis uncovers the essential physics behind a previously unrecognized universal wavy tail in the skyrmion spin texture. Notably, the resulting RKKY-like interaction between skyrmions is universal for all tilted skyrmions, irrespective of whether the titled easy-axis is from an external field or a crystalline magnetic anisotropy. These findings introduce a novel physical principle for the design of skyrmion molecules or skyrmion superstructures, which hold significant potential for applications in skyrmion-based spintronics and neuromorphic computing.
\end{abstract}
\maketitle

\textit{Introduction.--} Magnetic skyrmions are localized spin structures that are topologically non-trivial \cite{Bogdanov2001}. 
Since their discovery \cite{muhlbauer2009,Yu2011, Woo2016, Boulle2016, Kezsmarki2015, Pfleiderer1}, significant research 
has explored their fascinating physics and potential applications. This includes studies on skyrmion creation and annihilation
\cite{roadmap, Rossler2006, Tchoe2012,Nayak2017,jiang2015, Iwasaki2013, Hrabec2017, Jin2016}, morphology 
\cite{xs2018,stripe1,stripe2, paper1, paper2, paper3, paper4, paper5, paper6, paper7}, dynamics under various stimuli 
\cite{gong_prb_2020, Yuan2018, Schutte2014, Pinning1,Pinning2,response1, Pfleiderer3,ong,the2}, skyrmion-skyrmion 
interactions \cite{repel1,repel2,repel3,repel4,repel5,repel6,attractive1,attractive2,attractive3,Tattract,Tattract1,thickact1,thickact2,fruskm1, fruskm3} 
as well as their possible usages in technology \cite{racetrack1, racetrack2,gates}.

It is well known that two isolated skyrmions in a vertically magnetized chiral film always repel each other \cite{repel1, repel2, repel3,repel4,repel5,repel6}. 
However, recent studies have found that skyrmion-skyrmion interaction can be attractive under the tilted external 
fields \cite{attractive1, attractive2, attractive3}, or through thermal frustration \cite{Tattract}, or by varying sample 
thickness \cite{thickact1,thickact2}, and or through frustrated exchange interactions \cite{Tattract1,fruskm1, fruskm3}. 
Despite of all progress, the underlying physics of these interactions remains poorly understood, and a clear understanding 
of the conditions under which two skyrmions in a chiral film can attract each other is still lacking.

In this Letter, we demonstrate that skyrmion–skyrmion interactions in in-plane magnetized chiral films, caused by either a tilted 
magnetic field or magnetic anisotropy, are intrinsically Ruderman-Kittel-Kasuya-Yosida-type (RKKY-type), periodically alternating 
between attraction and repulsion. 
The tail of an isolated in-plane skyrmion inherently possesses a periodic spin structure along the direction of the skyrmion's 
center spin, with the same period as that of the interactions. When two skyrmions are aligned along this periodic direction and 
separated by an integer multiple of the period, their wavy tails coincide, allowing them to share the same wavy tail, which 
results in a reduction in total energy and an attractive interaction. Conversely, at separations of half-integer multiples of the 
period, the two tails are out of phase, and significantly deformation is needed in order for two skyrmions to connect each other 
smoothly, leading to an increase in total energy and a repulsive force between them. These findings provide new insights into the nature 
of skyrmion–skyrmion interactions and open up avenues for engineering skyrmion-based devices with controllable collective behavior.

\textit{Model and methodology.--} A chiral magnetic film of thickness $d$ lying in the $xy$-plane is modeled by the magnetic energy $E$, 
which consists of several components: the exchange energy $E_\mathrm{ex} =\frac{Ad}{2} \int |\nabla \mathbf{m}|^2 \, \mathrm{d}^2\mathbf{x}$, 
where the exchange stiffness $A = 30\,\mathrm{pJ/m}$ is fixed and $\mathbf{m}$ is the direction of magnetization with a saturation 
magnetization $M_s = 1\,\mathrm{MA/m}$; the Dzyaloshinskii–Moriya interaction (DMI) $E_\mathrm{DM}= \frac{Dd}{2} \int\left[ \mathbf{m} 
\cdot (\nabla \times \mathbf{m}) \right] \, \mathrm{d}^2\mathbf{x}$ with DMI strength $D= 4\,\mathrm{mJ/m}^2$; the anisotropy 
energy $E_\mathrm{an} = d\int-\frac{K}{2} (m_y\sin\theta+m_z\cos\theta)^2\mathrm{d}^2\mathbf{x}$ where $K$ is the 
anisotropy constant along the easy axis in the $yz$-plane, determined by the angle $\theta$; the magnetic dipolar energy, which can be 
incorporated into the effective anisotropy term $\frac{\mu_0 M_s^2 m_z^2}{2}$, with $\mu_0$ being the vacuum permeability; and 
the Zeeman energy $E_\mathrm{Ze} = -\mu_0 M_{\rm s} d \int \mathbf{H} \cdot \mathbf{m} , \mathrm{d}^2\mathbf{x}$, where 
$\mathbf{H}$ is the external magnetic field. In this study, only $K$, $\theta$, and $\mathbf{H}$ vary to model different situations. 
Two cases are considered: the first is $\mathbf{H}$ controlled anisotropy, where $\theta=0$ and $K = 1.256 \,\mathrm{MJ/m}^3$ are 
set to cancel the shape anisotropy, making the easy axis unsymmetrical and preferably aligning spin with $\mathbf{H}$; most results reported in the 
main text are derived from this case. The second case involves setting $\mathbf{H} = 0$ and using $K$ and $\theta$ to control the 
magnetic anisotropy, resulting in a symmetrical axis. The results of the second case are primarily presented in the Supplementary Materials \cite{suppl}.

Magnetization dynamics is governed by the Landau–Lifshitz–Gilbert (LLG) equation:
\begin{equation}
	\frac{\partial \mathbf{m}}{\partial t}
	= -\gamma \, \mathbf{m} \times \mathbf{H}_{\rm eff}
	+ \alpha \, \mathbf{m} \times \frac{\partial \mathbf{m}}{\partial t},
	\label{LLG eq}
\end{equation}
where $\gamma$ is the gyromagnetic ratio and $\alpha$ is the Gilbert damping. The effective field is 
$\mathbf{H}_{\rm eff}=-\delta E/\delta \mathbf{m}=A/(\mu_0 M_s) \nabla^2 \mathbf{m}+ K(m_z\cos\theta+m_y\sin\theta)(\cos\theta\mathbf{z}+\sin\theta\mathbf{y})-
D /(\mu_0 M_s)\nabla\times \mathbf{m}+ \mathbf{H}$. 
To obtain the total energy and stable spin structures of two skyrmions with a given displacement, we use 
Mumax3 package~\cite{MuMax3} to numerically solve the LLG equation on films of size 
$600\,\mathrm{nm} \times 600\,\mathrm{nm} \times 0.5\,\mathrm{nm}$, with proper constrains and the 
periodic boundary conditions in both $x$ and $y$ directions and lateral mesh size of  $1\,\mathrm{nm}$. 
\par 

\begin{figure}[htbp]
	\centering
	\includegraphics[width=0.48\textwidth]{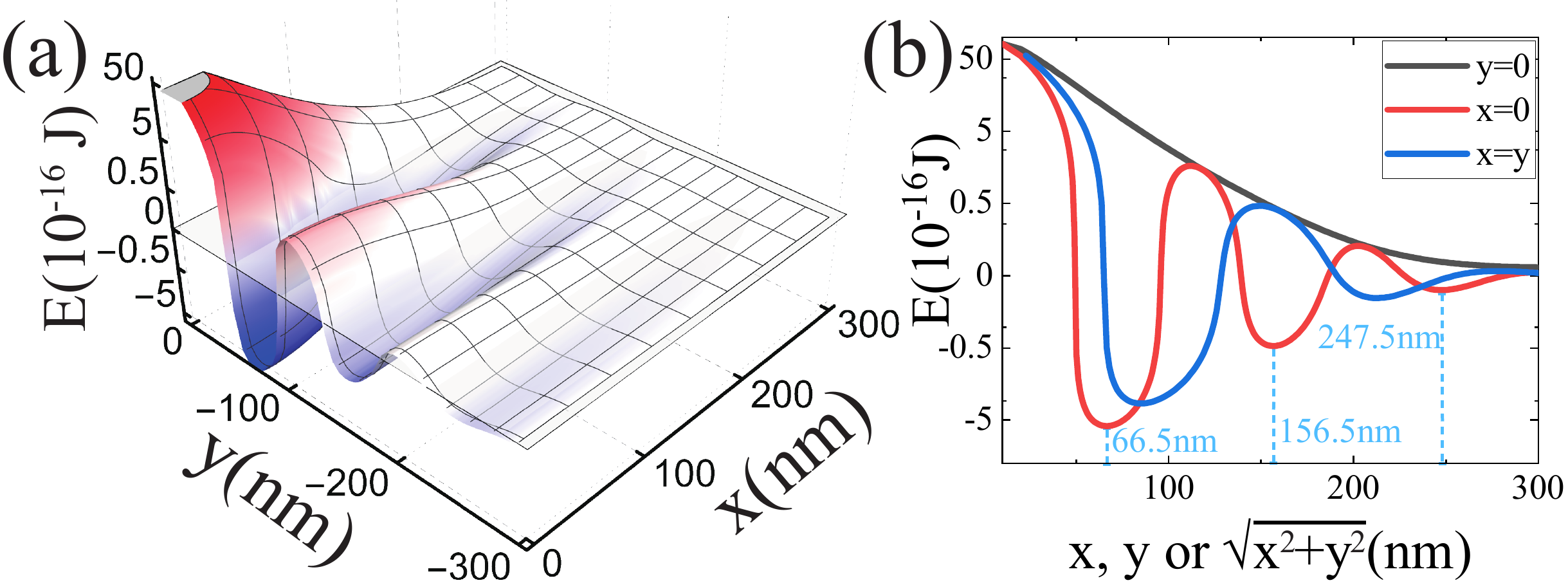}\\
	\caption{(a) Energy surface $E(x,y)$ of two skyrmions for $H=0.164$T along the $y$-axis. 
$E=0$ is chosen for two isolated skyrmions infinitely apart. (b) $E(x,y)$ as a function of  $x$ 
along $y = 0$ (black line), or $y$ along $x=0$ (red line), or $\sqrt{x^2+y^2}$ along $y=x$ (blue line). 
The vertical blue dash-lines indicate local minima. }
	\label{fig1}
\end{figure}

\textit{Skyrmion-skyrmion interaction.--} The nature of skyrmion interaction is given by the displacement 
dependence of potential energy. Thus, we fix the center of one skyrmion at the sample center $(0,0)$ and the 
other at $(x,y)$, with their center spin direction set to $\mathbf{m}=(0,1,0)$ in the background of spin states 
of $\mathbf{m}=(0,-1,0)$ (far from two skyrmion centers).  The remaining spins in the film follow the 
Landau-Lifshitz-Gilbert (LLG) equation and settle into a stable configuration. Figure~\ref{fig1}(a) is the 
energy surface $E(x,y)$ of the two skyrmions in their stable configuration when an external field of 
$H=0.164\,$T is applied along the $-y$-direction. Notably, $E(\pm x, \pm y)=E(x,y)$, so  it is sufficient to 
plot $E(x,y)$ in just one quadrant. The energy function $E(x,y)$ is clearly anisotropic, and the interaction 
between the two skyrmions is also anisotropic.  Along the $x$-axis ($y=0$), the total energy decreases 
monotonically with $x$ (black line in Fig. \ref{fig1}(b)), indicating that the two skyrmions repel each other. 
In contrast, along the $y$-axis ($x=0$), the energy profile exhibits RKKY-like oscillatory behaviour 
with multiple energy minima at $y_{min}=66.5, 156.5, 247.5\,$nm (red line in Fig. \ref{fig1}(b)), implying an alternating attractive 
and repulsive interactions between the two skyrmions. In other directions, oscillatory behaviour persists with the period of 
$\lambda /\sin \phi$, where $\phi$ is the angle between the skyrmion-skyrmion displacement $(x,y)$ and $y$-axis, $\lambda \approx 
156.5-66.5 \approx 247.5-156.5\approx 90\sim 91\,$nm which is not sensitive to $H$ and $\phi$. The blue curve in Fig. \ref{fig1}(b) 
is the energy variation along $y=x$ direction. According to energy surface $E(x,y)$, the skyrmion centred at $(x,y)$ will relax to its 
local minimum at $(0,y_{min})$ when the skyrmion is allowed to move. Which $y_{min}$ will end up with depends on the initial location 
of the skyrmion. Our simulations have confirmed this expectation. 

\begin{figure}[htbp]
	\centering
	\includegraphics[width=0.45\textwidth]{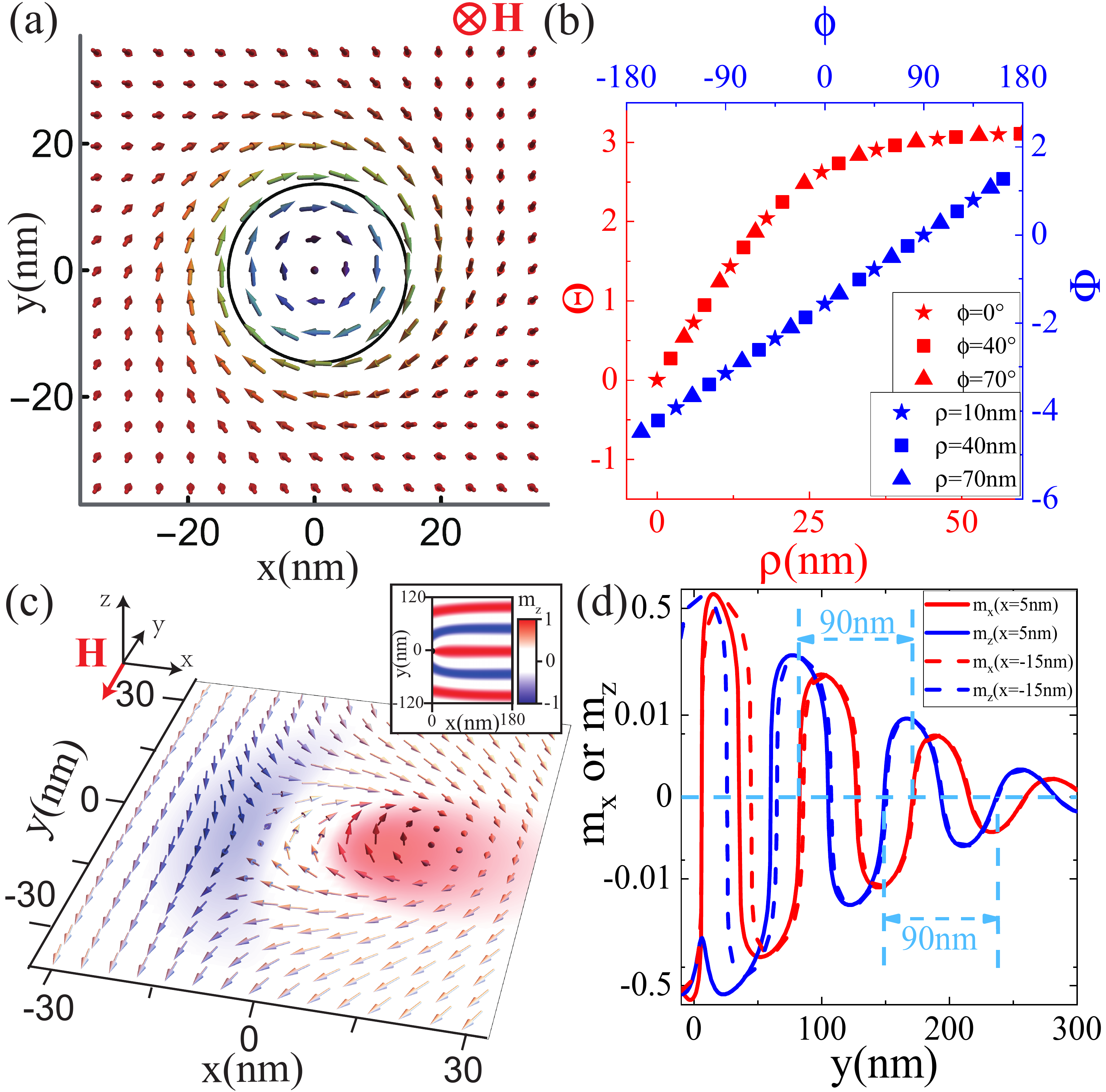}\\
	\caption{(a) A perpendicular skyrmion with $H = 0.164$T along the $-z$-axis. (b) $\Theta(\rho)$ (red) 
	for $\phi=0$ (stars),  $40^0$ (squares), $70^0$ (triangles) and $\Phi(\phi)$ (blue) for $\rho=10$nm 
	(stars), $40$nm (squares), $70$nm (triangles). (c) An in-plane skyrmion $H=0.164$T along the 
	$-y$-axis. Inset: density plots of $m_z>0$ (red) and $m_z<0$ (blue). The skyrmion center is indicated 
	by a dashed circle. (d) $m_x$ (red) and $m_z$ (blue) as a function of $y$ for $x=5$nm (solid curve) 
	and -15 nm (dashed lines).
}
	\label{fig2} 
\end{figure}

\textit{Periodic tail structure.--} To reveal the origin of the oscillatory attraction-repulsion interaction, we compare the spin structure 
of an isolated perpendicular skyrmion with that of an in-plane skyrmion. Figure~\ref{fig2}(a) shows that spin structure of a perpendicular 
skyrmion is isotropic, with spins varying monotonically and smoothly from up at the skyrmion center to down away from the center. 
In polar coordinates $(\rho,\phi)$, the polar and azimuthal angles $\Theta$ and $\Phi$ of $\mathbf{m}$ are functions of  
$\rho$ and $\phi$. Figure~\ref{fig2}(b) presents $\Theta(\rho)$ (red) for $\phi=0$ (stars), $40^0$ (squares), $70^0$ (triangles) and 
$\Phi(\phi)$ (blue) for $\rho=10$nm (stars), $40$nm (squares), $70$nm (triangles). The overlap of $\Theta(\rho)$ and $\Phi(\phi)$ 
demonstrate the isotropic spin structure. In contrast, the spin structure of an in-plane skyrmion (when the external field is along the 
$-y$-direction) is anisotropic as plotted in Fig.\ref{fig2}(c). 
The skyrmion has a stripy spin structure parallel to the $x$-axis with a well-defined period as shown in the inset of Fig.\ref{fig2}(c): 
red for $m_x>0$ and blue for $m_x<0$. The stripy structure can also be seen from $m_x(x=\text{const},y)$ (the red curves) and 
$m_z(x=\text{const},y)$ (the blue curves), exhibiting oscillatory behaviour as shown in Fig.\ref{fig2}(d). 

To see whether the periodical stripy tail of an in-plane skyrmion is intrinsic or not, we go back to the  
equations of stable spin textures~\cite{xs2018}, including in-plane skyrmions. For the easy-axis 
controlled by $\mathbf{H}$ in the $yz$-plane, $\Theta$ is defined as the angle between $\mathbf{m}$ 
and the easy-axis, and $\Phi$ is the azimuthal angle of $\mathbf{m}$ in the plane perpendicular to the 
easy-axis. In terms of nature length unit of $4A/\pi D$ and defining $\kappa' \equiv \pi^2 D^2/(16 
\mu_0 M_s A H)$,  equations for $\Theta$ and $\Phi$ are \cite{paper4,paper5,paper6},
\begin{equation} 
	\begin{aligned}
		2\nabla^2\Theta-\sin2\Theta\left(\nabla\Phi\right)^2-	\frac{4}{\pi}[\sin\theta_H\sin2\Theta\partial_y\Phi+&\\2\sin^2\Theta(\cos\Phi\partial_x\Phi+\cos\theta_H\sin\Phi\partial_y\Phi)]+\frac{2}{\kappa'}\sin\Theta&=0
		\label{EquTheta1}
	\end{aligned} 
\end{equation}

\begin{equation} 
	\begin{aligned}
		&-\sin2\Theta\left(\nabla\Theta\right)\cdot\left(\nabla\Phi\right)-\sin^2\Theta\nabla^2\Phi+\frac{2}{\pi}[\sin\theta_H\sin2\Theta\partial_y\Theta\\
		&+2\sin^2\Theta(\cos\Phi\partial_x\Theta+\cos\theta_H\sin\Phi\partial_y\Theta)]=0,
		\label{EquPhi1}
	\end{aligned}	
\end{equation}
where $\theta_H$ is the angle between external field $\mathbf{H}$ and the $z$-axis. For our in-plane skyrmion centred 
at $(0,0)$ under a filed of $0.164$T, $\theta_H=\pi/2$ and $\Theta(0,0) = 0, ~ \Theta(|x| \to \infty, or |y| \to \infty) = \pi$. 
Far from the skyrmion center, $|\pi-\Theta(x,y)| \ll 1$, and  Equations \eqref{EquTheta1} and \eqref{EquPhi1} become
\begin{equation} 
	\begin{aligned}
		\nabla^2\Theta + (\pi-\Theta)\left[\left(\nabla\Phi\right)^2-\frac{4}{\pi}\partial_y\Phi+
		\frac{1}{\kappa'}\right] = 0, 
		\label{EquTheta2}
	\end{aligned} 
\end{equation}
\begin{equation} 
	\begin{aligned}
		&2(\pi-\Theta)\left[\partial_x\Theta\partial_x\Phi +\partial_y\Theta \left(\partial_y\Phi-\frac{2}{\pi}\right) \right]\\&-(\pi-\Theta)^2(\nabla^2\Phi-2\cos\Phi\partial_x\Phi) = 0. 
		\label{EquPhi2}
	\end{aligned}	
\end{equation}
Importantly, \eqref{EquTheta2} and \eqref{EquPhi2} admit following asymptotic solution,
\begin{equation} 
	\begin{aligned}
		\Theta(x,y) =\pi- \pi e^{-k_x |x| - k_y |y|},~\Phi(y) = \frac{2}{\pi}y + \Phi_0, 
		\label{Theta}
	\end{aligned}	
\end{equation}
where $k_x^2 + k_y^2 = 1/\kappa' - 4/\pi^2$, and $\Phi_0$ is a constant. When $\kappa'< \pi^2/4$, $k_x^2+k_y^2>0$, the 
skyrmion is metastable (otherwise unstable), and the solution has a periodic wavy structure similar to conical state \cite{Pfleiderer1}.
Although $k_x^{-1}$ and $k_y^{-1}$ depend on model parameter $\kappa'$ (or $\kappa\equiv \pi^2 D^2/(16 A K)$ in 
the absence of a magnetic field and in an in-plane magnetized film, see Supplementary Materials 
\cite{suppl}), period $\lambda$ of $\Phi$ (the wavy tail) depends only on the length scale $4A/\pi D$  as 
$\lambda=2\pi \left(\frac{2}{\pi}\right) ^{-1}\times\frac{4A}{\pi D}=4\pi \frac{A}{D}\approx 93\,$nm,
not sensitive to the field strength (or magnetic anisotropy)! The approximate value agrees well with the 
simulated value of $90\sim 91$nm from Mumax3. For a perpendicular skyrmion under a 
perpendicular field, similar analysis does not have a wavy tail structure. 

\textit{Attractive interaction from the wavy tail.}-- 
We can now understand how two skyrmions with wavy tails give rise to an anisotropic and oscillatory 
interaction of well-defined period when the distance between two skyrmions varies. 
Along the $y$ direction, let us consider one skyrmion centred at $(0,0)$ and the other 
centred at $(0, y)$, obviously, two tails in the overlapped region are the same when 
$y= n\lambda$, and two skyrmions can share the same tail and save the formation energy of 
one such tail. This energy reduction can bind two skyrmions together, similar to the formation 
of a molecule from atoms. This energy deduction can be estimate from skyrmion profile of 
Eq. \eqref{Theta} as, 
\begin{equation}
	\begin{aligned} 
			\frac{\Delta E}{Ad} =&\int_{n\lambda/2}^{\infty}\mathrm{d}\mathbf{x}\int_{-\infty}^{\infty}\mathrm{d}\mathbf{y}\bigg[\left(\nabla\Theta\right)^2+
	\Theta^2\left(\nabla\Phi\right)^2-\\ & -\frac{4\Theta^2}{\pi}\partial_y\Phi+\frac{\Theta^2}{\kappa'}+\Theta^2\cos\Phi\partial_x\Theta\bigg] 
		\\ =&\left(\frac{4(k_x^2+k_y^2)}{k_x k_y}-
		\frac{4k_y}{\pi k_y^2+\frac{4}{\pi}}\right)e^{-k_y n\lambda}>0. 
	\end{aligned}
\end{equation} 
In the case of $ y=(n+1/2)\lambda$, the two wavy tails are out of phase. To smoothly connecting 
two skyrmions together, both skyrmions must deform their structures that require an increase of 
the total energy, resulting in a repulsive force.

\begin{figure}[htbp]
	\centering
	\includegraphics[width=0.48\textwidth]{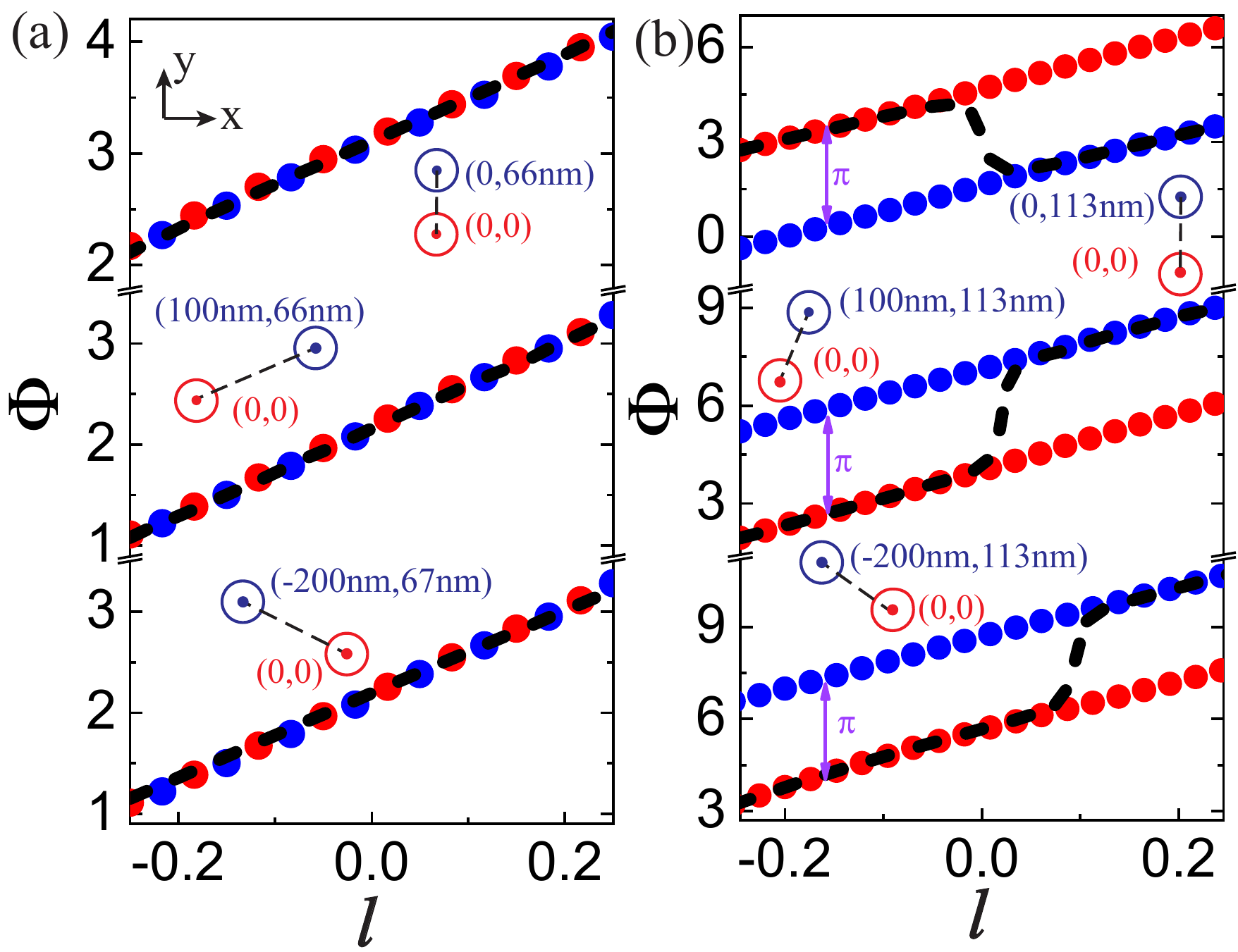}\\
	\caption{$\Phi(l)$ of skyrmion centred at $(0,0)$ (red symbols) and $(x_0, y_0)$ (blue 
symbols), and $\Phi(l)$ of the two coupled skyrmions (black dashed line) for $(x_0, y_0)=
(0,66\mathrm{nm} )$, $(100\mathrm{nm},66\mathrm{nm})$, and $(-200\mathrm{nm},
67\mathrm{nm})$ (a) and for $(x_0, y_0)=(0,113\mathrm{nm} )$, $(100\mathrm{nm},
113\mathrm{nm})$, and $(-200\mathrm{nm},113\mathrm{nm})$ (b). 
	}
	\label{fig3} 
\end{figure}

To further support our assertion that attraction comes from the match of two skyrmion tails in the 
overlapped region, we compare $\Phi$ in the overlapped region of two independent skyrmions 
at a fixed separation, as well as the $\Phi$ of the two skyrmions in its final relaxed structure.   
One skyrmion is centred at $(0,0)$ and the other at $(x_0, y_0)$. 
We first use Mumax3 to simulate only one skyrmion at $(0,0)$ with the model parameters specified above and 
computer $\Phi (l)$ [red symbols in Figs. \ref{fig3}(a)-(b)], where $l\equiv |\sqrt{x_0^2+
y_0^2}(|y|/|y_0|-1/2)$ measures distance from the middle point between the two skyrmion 
along the line of $y=(y_0/x_0)x$. Then, we simulate only one skyrmion 
centred at $(x_0, y_0)$ for its $\Phi (l)$ [blue symbols in Figs. \ref{fig3}(a)-(b)]. 
Finally, we simulate a bi-skyrmion molecule, or two bonded skyrmions respectively at 
$(0,0)$ and $(x_0, y_0)$ and compute $\Phi (l)$ [black dashed lines in Figs. \ref{fig3}(a)-(b)]. 
Figure~\ref{fig3}(a) presents $\Phi (l)$ when two tails are in good match or two syrmions 
attract each other for $(x_0, y_0)=(0,66\mathrm{nm} )$, $(100\mathrm{nm}, 66\mathrm{nm})$, 
and $(-200\mathrm{nm},67\mathrm{nm})$. $\Phi (l)$ is linear in $l$ with the expected slop of 
$\frac{D}{2A} =0.0667\,\mathrm{nm}^{-1}$. The difference in $\Phi (l)$ among two independent 
skyrmions and the bi-skyrmion molecule can hardly been seen, showing small deformation of 
the bi-skyrmion molecule. The period of oscillatory stripy tail along direction of $(x_0, y_0)$ is 
$\lambda/\cos(\phi)$ where $\phi=tan^{-1}(y_0/x_0)$. $(x_0, y_0)=(0,66\mathrm{nm} )$ is 
slightly different from the first minimum of $(0,66.5\mathrm{nm})$ due to a small deformation 
of two skyrmions from the isolated ones. 

In the opposite case when two skyrmions repel each other, their tails in the overlapped region 
are out of phase, or the difference between two $\Phi$'s is $\pi$ as shown in Fig. ~\ref{fig3}(b) 
for $(x_0, y_0)=(0,113\mathrm{nm} )$, $(100\mathrm{nm},113\mathrm{nm})$, and 
$(-200\mathrm{nm},113\mathrm{nm})$ which are close to the peaks of total energy in 
Fig. ~\ref{fig1}(a). For all three cases,  $\Phi (l)$ of the skyrmion centred at $(0,0)$ 
differs from $\Phi (l)$ of the skyrmion centred at  $(x_0, y_0)$ by $\pi$ for all $l$. 
Furthermore, the same as those in Fig. ~\ref{fig3}(a),  $\Phi (l)$ is linear with the slop of 
$\frac{D}{2A}$. Because of the mismatch of the two tails, spin structures of two coupled 
skyrmions in the overlapped region must greatly deform from their original structures. 
$\Phi (l)$ of two relaxed skyrmion should be different from the value of one skyrmion.
Indeed, as shown in Fig.~\ref{fig3}(b), $\Phi (l)$ of fully relaxed spin structure of two 
coupled skyrmions centred at $(0,0)$ and $(x_0, y_0)$, respectively, are not only different from 
both red and blue symbols, but also not linear any more. 

\begin{figure}[htbp]
	\centering
	\includegraphics[width=9cm]{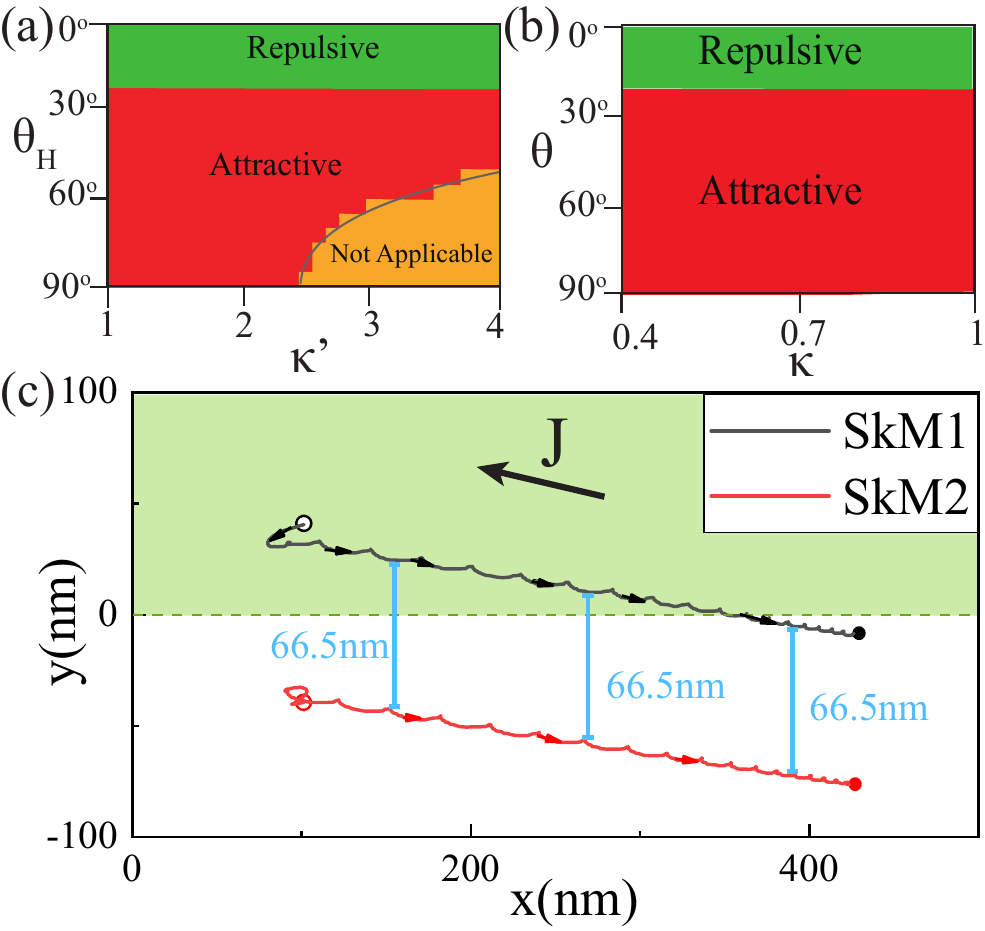}\\
	\caption{Phase diagrams of RKKY-like interaction (the red), repulsive interaction (the green) , and 
non-existence of isolated skyrmion (the orange) in $\kappa'-\theta_H$ plane for $\mathbf{H}$ controlled asymmetric 
anisotropy (a) and in $\kappa-\theta$ planes for $K-\theta$ controlled symmetric magnetic anisotropy (b).  
(c) Trajectories of a bi-skyrmion molecule under a spin–orbit torque. Torque is applied only on the top skyrmion whose motion drags the other skyrmion to move together.}
	\label{fig4}
\end{figure}

\textit{Phase diagram--}
We have seen that the interactions between two perpendicular skyrmions and between two in-plane 
skyrmions are fundamentally different. A natural question is whether the oscillatory skyrmion-skyrmion 
interaction exists in between. Namely, is skyrmion-skyrmion interaction RKKY-like if the easy axis is in 
an arbitrary direction in the $yz$-plane? This question can be easily investigated by changing $\theta_H$. 
According to Eqs. \eqref{EquTheta1} and \eqref{EquPhi1}, stable spin textures are fully 
determined by $\theta_H$ and $\kappa'$. Our Mumax3 simulations lead to the phase diagram of  
Figs.~\ref{fig4}(a) in the $\kappa'-\theta_H$ plane that distinguish three phases: 
RKKY-like skyrmion-skyrmion interaction (the red); repulsive skyrmion-skyrmion interaction (the green); 
and region without isolated skyrmion (the orange). Interestingly, within the model parameters used in 
this study, RKKY-like interaction appears only when $\theta_H>23^\circ$. From the analysis of 
Eqs. \eqref{EquTheta1} and \eqref{EquPhi1}, boundary between the red and orange 
is given by $\kappa'=\left[\pi \sin(\theta_H)\right]^2$ (grey curve) which agree excellently with 
the simulations ( Supplementary Materials \cite{suppl}). 

Although results presented so far are all from asymmetrical anisotropy due to $\mathbf{H}$. 
A natural question is whether they are also true to the case of usual symmetric crystalline magnetic anisotropy. 
We have carried out similar studies in the absence of $\mathbf{H}=0$ and a proper $K$ to address the issue by varying $\theta$. 
Similar equations like Eqs. \eqref{EquTheta1} and \eqref{EquPhi1} can be obtained which depend only 
on $\theta$ and $\kappa\equiv (\pi^2D^2)/(16AK)$, i.e. stable spin textures are uniquely determined 
by these two parameters. Figure~\ref{fig4}(b) is the phase diagram in the $\kappa-\theta$ plane. Interestingly and surprisingly, 
RKKY-like interaction again appears only for $\theta>23^\circ$ and $\kappa < 1$ for isolated skyrmion, 
in agreement with previous results~\cite{paper1,paper2,paper7}.
The details results are presented in Supplementary Materials \cite{suppl}. 
Furthermore, the physics presented here does not change if one uses interfacial DMI (see Supplementary 
Materials \cite{suppl}), rather than the bulk DMI present above. As expected, results presented here do not 
depend on the origins of the anisotropy, nature of the anisotropy--symmetric or asymmetric, 
or the type of DMI interaction.

With tunable skyrmion-skyrmion interactions, one can design more complicated skyrmion superstructures. 
Below, we demonstrate that a bi-skyrmion molecule binding together by the attraction moves as one object 
under a spin-orbit torque acting only on one part of the molecule. Figure~\ref{fig4}(c) plots the trajectories 
of the molecule consisting two skyrmions located on opposite sides of $y=0$ and separated by a distance 
of $66.5\,\mathrm{nm}$. A spin–orbit torque from a current density of $10^{11}\,\mathrm{A/m^2}$ 
acting only on the skyrmion in $y > 0$ region. Notably, two skyrmions move together and maintain their 
bond length of $66.5\,\mathrm{nm}$ during the whole process.

\textit{Conclusions.--}  
In conclusion, skyrmions in a tilted magnetic field or in a title magnetic anisotropy with a sufficient large 
component in the film plane are anisotropic and have periodically wavy tails. This wavy tail leads to 
skyrmion–skyrmion attraction when the two tails match with each other in the overlapped region, and repulsion 
when their tails mismatch with each other. As a result, two tilted skyrmions exhibit RKKY-like interaction, 
oscillating between attraction and repulsion with a period that is determined by the period of wavy tail. 
The period is proportional to $A/D$ and inversely proportional to $\sin\theta_H$ (for field-controlled anisotropy) 
or $\sin\theta$ (for crystalline anisotropy). The direction of periodical tail structure is determined by the type 
of DMI interaction and easy axis direction (see Supplementary Materials~\cite{suppl}). 

Anisotropic and oscillatory skyrmion–skyrmion interaction gives rise to new possibilities in skyrmion-based spintronics. 
The presence of multiple optimal skyrmion–skyrmion separations offers an additional degree of freedom for 
controlling the dynamics of isolated skyrmions. Our findings thus open a promising avenue for advanced skyrmion 
manipulation such as creating skyrmion molecules or more complicated skyrmion superstructures. 

\begin{acknowledgments} 
\textit{Acknowledgments--}This work is supported by the University Development 
Fund of the Chinese University of Hong Kong, Shenzhen, and NSFC Grant (No. 12374122). H. Y. Yuan is supported by the National Key R$\&$D Program of China (2022YFA1402700) and NSFC Grant (No. 12574132).
\end{acknowledgments} 
%
%
%
%


\begin{thebibliography}{99}
	


\bibitem{Bogdanov2001}A. N. Bogdanov and U. K. R\"{o}{\ss}ler,
\href{https://journals.aps.org/prl/abstract/10.1103/PhysRevLett.87.037203}
{Phys. Rev. Lett., {\bf 87}, 037203 (2001)}.

\bibitem{muhlbauer2009}S. M\"{u}hlbauer, B. Binz, F. Jonietz, 
C. Pfleiderer, A. Rosch, A. Neubauer, R. Georgii1, and P. B\"{o}ni,
\href{https://science.sciencemag.org/content/323/5916/915.abstract}
{Science {\bf 323}, 915 (2009)}.

\bibitem{Yu2011}X. Z. Yu, N. Kanazawa, Y. Onose, K. Kimoto, W. Z. Zhang, 
S. Ishiwata, Y. Matsui, and Y. Tokura,
\href{https://www.nature.com/articles/nmat2916}
{Nat. Mat. {\bf 10}, 106 (2011)}.


\bibitem{Woo2016}S. Woo, K. Litzius, B. Kr\"{u}ger, M.Y. Im, L. Caretta,
K. Richter, M. Mann, A. Krone, R. M. Reeve, M. Weigand, \textit{et al.}
\href{https://www.nature.com/articles/nmat4593}
{Nat. Mat. {\bf 15}, 501 (2016)}.
	
	
\bibitem{Boulle2016}O. Boulle, J. Vogel, H. Yang, S. Pizzini, 
D. de S. Chaves, A. Locatelli, T. O. Menteş, A. Sala, L. D. Buda-Prejbeanu, 
O. Klein, \textit{et al.}
\href{https://www.nature.com/articles/nnano.2015.315}
{Nat. Nano. {\bf 11}, 449 (2016)}.

	
\bibitem{Kezsmarki2015}I. K\'{e}zsm\'{a}rki, S. Bord\'{a}cs, P. Milde,
E. Neuber, L. M. Eng, J. S. White, H. M. R{\o}nnow, C. D. Dewhurst, 
M. Mochizuki, K. Yanai, \textit{et al.}
\href{https://www.nature.com/articles/nmat4402}
{Nat. Mat. \textbf{14}, 1116 (2015)}.


\bibitem{Pfleiderer1}W. Munzer, A. Neubauer, T. Adams, S. M\"{u}hlbauer, 
C. Franz, F. Jonietz, R. Georgii,  P. B\"{o}ni, B. Pedersen, M. Schmidt, 
A. Rosch, and C. Pfleiderer,  
\href{https://journals.aps.org/prb/abstract/10.1103/PhysRevB.81.041203}
{Phys. Rev. B {\bf 81}, 041203 (2010)}.		


\bibitem{roadmap}C. Back, V. Cros, H. Ebert, K. Everschor-Sitte, 
A. Fert, M. Garst, T. P. Ma, S. Mankovsky, T. L. Monchesky, 
M. Mostovoy, \textit{et al.}
\href{https://iopscience.iop.org/article/10.1088/1361-6463/ab8418/meta}
{J. Phys. D {\bf 53}, 363001 (2020)}.

\bibitem{Rossler2006}U. K. R\"{o}{\ss}ler, A. N. Bogdanov, and C. Pfleiderer, 
\href{https://www.nature.com/articles/nature05056}
{Nature {\bf 442}, 797 (2006)}.


\bibitem{Tchoe2012}Y. Tchoe and J. H. Han,
\href{https://journals.aps.org/prb/abstract/10.1103/PhysRevB.85.174416}
{Phys. Rev. B {\bf 85}, 174416 (2012)}.


\bibitem{Nayak2017}A. K. Nayak, V. Kumar, T. Ma, P. Werner, E. Pippel,  
R. Sahoo, F. Damay, U.K. R\"{o}{\ss}ler, C. Felser, and S. S. P. Parkin,  
\href{https://www.nature.com/articles/nature23466}
{Nature \textbf{548}, 561 (2017)}.


\bibitem{jiang2015}W. Jiang, P. Upadhyaya, W. Zhang, G. Yu, M. B. Jungfleisch, 
F. Y. Fradin, 
J. E. Pearson, Y. Tserkovnyak, K. L. Wang, O. Heinonen, \textit{et al.}
\href{https://www.science.org/doi/full/10.1126/science.aaa1442}
{Science {\bf 93}, 283 (2015)}.


\bibitem{Iwasaki2013}J. Iwasaki, M. Mochizuki, and N. Nagaosa.
\href{https://www.nature.com/articles/ncomms2442}
{Nat. Comm. {\bf 4}, 1 (2013)}.



\bibitem{Hrabec2017}A. Hrabec, J. Sampaio, M. Belmeguenai, I. Gross, 
R. Weil, S. M. Chérif, A. Stashkevich, V. Jacques, A. Thiaville, and S. Rohart,
\href{https://www.nature.com/articles/ncomms15765}
{Nat. Comm. {\bf 8}, 1 (2017)}.
	
	
\bibitem{Jin2016}C. Jin, C. Song, J. Wang, and  Q. Liu
\href{https://pubs.aip.org/aip/apl/article/109/18/182404/31932/Dynamics-of-antiferromagnetic-skyrmion-driven-by}
{Appl. Phys. Lett. {\bf 09}, 182404 (2016)}.
	
	

	
\bibitem{stripe1}Y. Z. Wu, C. Won, A. Scholl, A. Doran, H. W. Zhao, 
X. F. Jin, and Z. Q. Qiu,
\href{https://journals.aps.org/prl/abstract/10.1103/PhysRevLett.93.117205}
{Phys. Rev. Lett.  {\bf 93}, 117205 (2004)}.
	
\bibitem{stripe2}J. Dho, Y. N. Kim, Y. S. Hwang, J. C. Kim, and N. H. Hur,
\href{https://journals.aps.org/prl/abstract/10.1103/PhysRevLett.93.117205}
{Phys. Rev., B Condens. Matter {\bf 336}, 136 (2003)}.
	
\bibitem{paper1}X. R. Wang, X. C. Hu, and H. T. Wu,
\href{https://www.nature.com/articles/s42005-021-00646-9}	
{Commun. Phys. {\bf 4}, 142 (2021)}.
	
\bibitem{paper2}H. T. Wu, X. C. Hu, K. Y. Jing, and X. R. Wang,
\href{https://www.nature.com/articles/s42005-021-00716-y}
{Commun. Phys. {\bf 4}, 210 (2021)}.
	
\bibitem{paper3}H. T. Wu, X. C. Hu, and X. R. Wang, 
\href{https://link.springer.com/article/10.1007/s11433-021-1852-8}
{Sci. China- Phys. Mech. Astron. {\bf 65}, 247512 (2022)}.
	
\bibitem{paper4}X. C. Hu, H. T. Wu, and X. R. Wang,
\href{https://pubs.rsc.org/en/content/articlehtml/2022/nr/d2nr01300b}
{Nanoscale, {\bf 14}, 7516 (2022)}.
	
	
\bibitem{paper5}X. R. Wang and X. C. Hu,
\href{https://journals.aps.org/prb/abstract/10.1103/PhysRevB.107.174412}
{Phys. Rev. B, {\bf 107}, 174412 (2023)}.
	
	
\bibitem{paper6}X. R. Wang, X. C. Hu, and Z. Z. Sun,
\href{https://pubs.rsc.org/en/content/articlehtml/2022/nr/d2nr01300b}
{Nano Lett.  {\bf 23}, 3954 (2023)}.
	

\bibitem{paper7}M. V. Wijethunga, X. C. Hu, and X. R. Wang,
\href{https://www.nature.com/articles/s42005-025-01980-y}
{Commun. Phys. {\bf 8}, 123 (2025)}.
	
	
\bibitem{xs2018} X. S. Wang, H. Y. Yuan, and X. R. Wang,
\href{https://www.nature.com/articles/s42005-018-0029-0}
{Commun. Phys. {\bf 1}, 1 (2018)}.


\bibitem{Schutte2014}C. Sch\"{u}tte, J. Iwasaki, A. Rosch, and N. Nagaosa,
\href{https://journals.aps.org/prb/abstract/10.1103/PhysRevB.90.174434}
{Phys. Rev. B {\bf 90}, 174434 (2014)}.


\bibitem{Yuan2018}H. Y. Yuan, X. S. Wang, M. H. Yung, and X. R. Wang, 
\href{https://link.aps.org/doi/10.1103/PhysRevB.99.014428}
{Phys. Rev. B \textbf{99}, 014428 (2019)}.


\bibitem{gong_prb_2020}X. Gong, H. Y. Yuan, and X. R. Wang,
\href{https://journals.aps.org/prb/abstract/10.1103/PhysRevB.101.064421}
{Phys. Rev. B {\bf 101}, 064421 (2020)}.	


\bibitem{response1}X. Z. Yu, N. Kanazawa, W. Z. Zhang, T. Nagai,
T. Hara, T. Kimoto, K. Matsui, Y. Onose, and Y. Tokura, 
\href{https://journals.aps.org/prl/abstract/10.1103/PhysRevLett.96.207202}
{Nat. Comm. {\bf 3}, 1 (2012)}.


\bibitem{Pfleiderer3}A. Neubauer, C. Pfleiderer, B. Binz, A. Rosch, R. Ritz, 
P. G. Niklowitz, and P. B\"{o}ni,  
\href{https://journals.aps.org/prl/abstract/10.1103/PhysRevLett.102.186602}
{Phys. Rev. Lett. {\bf 102}, 186602 (2009)}.


\bibitem{ong}M. Lee, W. Kang, Y. Onose, Y. Tokura, and  N. P. Ong,
\href{https://journals.aps.org/prl/abstract/10.1103/PhysRevLett.102.186601}
{Phys. Rev. Lett. {\bf 102}, 186601 (2009)}.


\bibitem{Pinning1}X. Gong, K. Y. Jing, J. Lu, and X. R. Wang,
\href{https://journals.aps.org/prb/abstract/10.1103/PhysRevB.105.094437}
{Phys. Rev. B {\bf 105}, 094437 (2022)}.


\bibitem{Pinning2}S. Z. Lin, C. Reichhardt, C. D. Batista, and A. Saxena,
\href{https://journals.aps.org/prb/abstract/10.1103/PhysRevB.87.214419}
{Phys. Rev. B {\bf 87}, 214419 (2013)}.


\bibitem{the2}J. Zang; M. Mostovoy, J. H. Han, and N. Nagaosa, 
\href{https://journals.aps.org/prl/abstract/10.1103/PhysRevLett.107.136804} 
{Phys. Rev. Lett. {\bf 107}, 136804 (2011)}.

	
	
\bibitem{repel1}X. Zhang, G. P. Zhao, H. Fangohr, J. P. Liu, W. X. Xia, J. Xia,
and F. J. Morvan, 
\href{https://www.nature.com/articles/srep07643}
{Sci. Rep. {\bf 5}, 1 (2015)}.
	
	
\bibitem{repel2}S. Z. Lin, C. Reichhardt, C. D. Batista, and A. Saxena,
\href{https://journals.aps.org/prb/abstract/10.1103/PhysRevB.87.214419}
{Phys. Rev. B {\bf 87}, 214419 (2013)}.
	
	
\bibitem{repel3} R. Brearton, G. van der Laan, and T. Hesjedal,
\href{https://journals.aps.org/prb/abstract/10.1103/PhysRevB.101.134422}
{Phys. Rev. B {\bf 101}, 134422(2020)}.

\bibitem{repel4} D. Capic, D. A. Garanin, and E. M. Chudnovsky, 
\href{https://iopscience.iop.org/article/10.1088/1361-648X/ab9bc8/meta}
{J. Condens. Matter Phys. {\bf 32}, 415803(2020)}.
	
\bibitem{repel5}Y. Wang, J. Wang, T. Kitamura, H. Hirakata1, and  T. Shimada,
\href{https://iopscience.iop.org/article/10.1088/1361-648X/ab9bc8/meta}
{Phys. Rev. Appl. {\bf 18}, 044024 (2022)}.

\bibitem{repel6}S. Z. Lin and S. Hayami,
\href{https://journals.aps.org/prb/abstract/10.1103/PhysRevB.93.064430}
{Phys. Rev. B {\bf 93}, 064430(2016)}.


\bibitem{attractive1}S. Z. Lin and A. Saxena,
\href{https://journals.aps.org/prb/abstract/10.1103/PhysRevB.92.180401}
{Phys. Rev. B, {\bf 92}, 180401 (2015)}.


\bibitem{attractive2}M. Kameda, R. Koyama, T. Nakajima, and Y. Kawaguchi,
\href{https://journals.aps.org/prb/abstract/10.1103/PhysRevB.104.174446}
{Phys. Rev. B {\bf 104}, 174446 (2021)}.


\bibitem{attractive3} K. L. Tiwari, J. Lavoie, T. Pereg-Barnea, and W. A. Coish,
\href{https://journals.aps.org/prb/abstract/10.1103/PhysRevB.100.125414} 
{Phys. Rev. B {\bf 100}, 125414(2019)}.	


\bibitem{Tattract}A. O. Leonov and U. K. R\"{o}ßler,
\href{https://www.mdpi.com/2079-4991/13/5/891}
{Nanomaterials {\bf 13}, 891 (2023)}.

\bibitem{thickact1}J. C. Loudon, A. O. Leonov, A. N. Bogdanov, M. C. Hatnean, 
and G. Balakrishnan,
\href{https://journals.aps.org/prb/abstract/10.1103/PhysRevB.97.134403}
{Phys. Rev. B {\bf 97}, 134403 (2018)}.
	
	
\bibitem{thickact2}H. Du, X. Zhao, F. N. Rybakov, A. B. Borisov, S. Wang, J. Tang, 
C. Jin, C. Wang, W. Wei, N. S. Kiselev, \textit{et al.}
\href{https://journals.aps.org/prl/abstract/10.1103/PhysRevLett.120.197203}
{Phys. Rev. Lett. {\bf 120}, 197203 (2018)}.
	
\bibitem{Tattract1}L. R\'{o}zsa, A. De\'{a}k, E. Simon, R. Yanes, and
L. Udvardi,
\href{https://journals.aps.org/prl/abstract/10.1103/PhysRevLett.117.157205}
{Phys. Rev. Lett. {\bf 117}, 157205 (2016).}

\bibitem{fruskm1}L. R\'{o}zsa, A. De\'{a}k, E. Simon, R. Yanes, L. Udvardi, L. Szunyogh, and U. Nowak,
\href{https://journals.aps.org/prl/abstract/10.1103/PhysRevLett.117.157205}
{Phys. Rev. Lett. {\bf 117}, 157205 (2016).}

\bibitem{fruskm3}A. O. Leonov and M. Mostovoy,
\href{https://www.nature.com/articles/ncomms9275}
{Nat. Commun. {\bf 6}, 8275 (2015).}

\bibitem{racetrack1}R. Tomasello, E. Martinez, R. Zivieri, \textit{et al.}%
\href{https://link.springer.com/content/pdf/10.1038/srep06784.pdf}
{Sci. Rep. {\bf 4}, 1 (2014).}


\bibitem{racetrack2}W. Kang, Y. Huang, C. Zheng, W. Lv, N. Lei, Y. Zhang, 
X. Zhang, Y. Zhou, and W. Zhao,
\href{https://link.springer.com/content/pdf/10.1038/srep23164.pdf}
{Sci. Rep. {\bf 6}, 1 (2016).}

\bibitem{gates}S. Luo, M. Song, X. Li, Y. Zhang, J. Hong, X. Yang, X. Zou, 
N. Xu, and L. You,
\href{https://pubs.acs.org/doi/full/10.1021/acs.nanolett.7b04722}
{Nano Lett. {\bf 18}, 1180 (2018).}


\bibitem{MuMax3}A. Vansteenkiste, J. Leliaert, M. Dvornik, 
M. Helsen, F. Garcia-Sanchez, and B. V. Waeyenberge, 
\href{https://aip.scitation.org/doi/full/10.1063/1.4899186}
{AIP. Adv. {\bf 4}, 107133 (2014)}.
	
	
\bibitem{biskm1}B. G\"{o}bel, J. Henk, and I. Mertig,
\href{https://journals.aps.org/prl/abstract/10.1103/PhysRevLett.120.197203}
{Sci. Rep. {\bf 9}, 9521 (2019).}
	
	
\bibitem{biskm2}X. Z. Yu, Y. Tokunaga, Y. Kaneko, W.Z. Zhang2 K. Kimoto, 
Y. Matsui, Y. Taguchi, and Y. Tokura,
\href{https://www.nature.com/articles/ncomms4198.pdf}
{Nat. Commun. {\bf 5}, 3198 (2014).}
	
	

\bibitem{suppl}See Supplementary Material at 
\href{https://}
for wavy tail with tilted field, anisotropy with tilted easy axis and 
DMI type, includes Ref.[24, 25].

	
\end{thebibliography}
\end{document}


\title{Supplemental Materials for “RKKY-like Interactions Between Two Magnetic Skyrmions"}

\author{X. C. Hu}

\affiliation{School of Science and Engineering, Chinese University of Hong Kong (Shenzhen), Shenzhen, 51817, China}
\author{H. Y. Yuan}
\affiliation{Institute for Advanced Study in Physics, Zhejiang University, 310027 Hangzhou, China}
\author{X. R. Wang}
\email[Corresponding author: ]{phxwan@cuhk.edu.cn}
\affiliation{School of Science and Engineering, Chinese University of Hong Kong (Shenzhen), Shenzhen, 51817, China}
\date{\today}

\maketitle

\section{Skyrmion-skyrmion interaction in chiral films with titled symmetric magnetic anisotropy}

In the absence of external field $H = 0$ and with a titled magnetic anisotropy of $E_\mathrm{an} = d\iint-[\frac{K}{2} (m_y\sin\theta+m_z\cos\theta)^2+\frac{K_1}{2}m_z^2]\mathrm{dxdy}$, where $K_1 = 1.256 \,\mathrm
{MJ/m}^3$, $K=0.365\mathrm{MJ/m}^3$, and the rest model parameters are the same as those used in the main text, 
we repeat what have been done in the main text. As mentioned in the main text, $K_1$ cancels the shape anisotropy such 
that $K$ is the net symmetric magnetic anisotropy in the $yz$-plane with a titled angle of $\theta$ from the $z$-axis.  
These parameters correspond to $\kappa\equiv\pi^2D^2/(16AK)=0.9<1$ such that isolated skyrmions are metastable 
states \cite{paper1,paper7}. Different from $H$-induced asymmetric magnetic anisotropy where a spin prefers 
aligning along the field direction, this anisotropy equally prefers spins in both directions of the easy axis. 
We consider again the total energy $E(x,y)$ of two skyrmions centred at $(0,0)$ and at $(x,y)$, respectively, in the 
background of ferromagnetic state of $\mathbf{m}=(0,-\sin\theta,-\cos\theta)$. The spins at the skyrmions centres 
are fixed along $(0,\sin\theta,\cos\theta)$. 

\begin{figure}[htbp]
	\centering
	\includegraphics[width=0.96\textwidth]{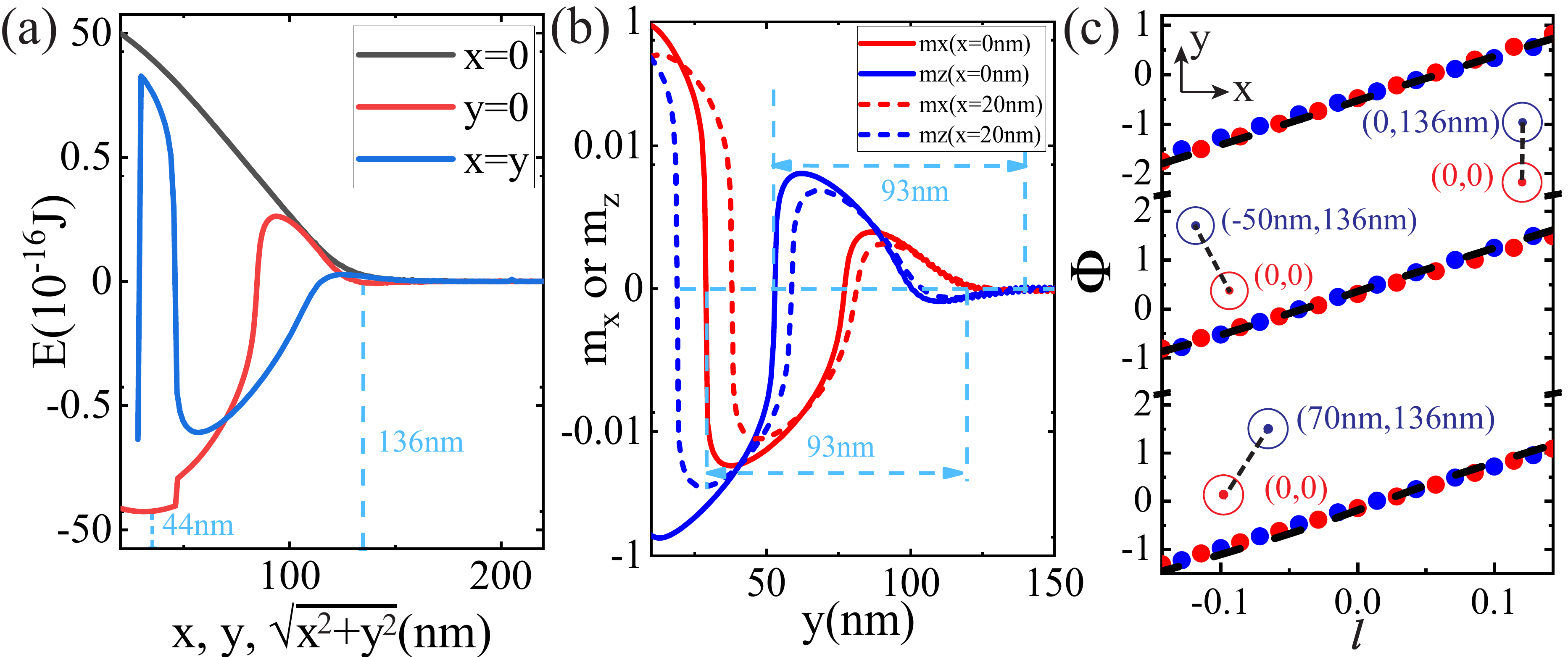}\\
	\caption{(a) An in-plane skyrmion with $K=0.365\,\mathrm{MJ/m^3}$ an $\theta=0$ 
	along the $-y$-axis. Inset: density plots of $m_x>0$ (red) and $m_x<0$ (blue). 
	The skyrmion center is denoted by a small circle. (b) $m_x$ (red) and $m_z$ (blue) as 
	a function of $y$ for $x=0$nm (solid curve) and 20 nm (dashed lines). (c) $E(x,y)$ as 
	a function of $x$ along $y = 0$ (black line), or $y$ along $x=0$ (red line), or 
	$\sqrt{x^2+y^2}$ along $y=x$ (blue line). The vertical blue dash-lines 
	indicate local minimum. (d) An in-plane skyrmion of $Q=2$ with same set of 
	parameters. (e) $\Phi(l)$ of skyrmion centred at $(0,0)$ (the black lines) and 
	$(x_0, y_0)$ (the red lines), and $\Phi(l)$ of the two skyrmions in their fully  
	relaxed state (the blue dotted lines) for  $(x_0, y_0)=(0,136\mathrm{nm} )$, 
	$(-50\mathrm{nm},136\mathrm{nm})$, and $(70\mathrm{nm},136\mathrm{nm})$  
	}
	\label{fig1}
\end{figure}

Figure \ref{fig1}(a) shows the total energy $E(x,y)$ for the case of $\theta=\pi/2$ (easy-axis along the $y$). 
The red, black, and blue curves are $E(x,y)$ for two skyrmions align along the $y$-axis, the $x$-axis, and along the line of $y=x$. Similar to the case in the main text, $E(x,y)$ monotonically decreases along the $x$-axis, or a repulsive interaction between two skyrmions in the direction. $E(x,y)$ is oscillatory along all other directions with 
a well-defined period $\lambda(x)=\lambda_0/\sin\phi$ where $\phi=\tan^{-1}(y/x)$, or demonstrating a 
RKKY-like interaction. Like that in the case of a titled field,  $\lambda_0\approx136-44\approx90\,$nm is 
not sensitive to value of $K$ and depends only on $D/A$ and $\theta$. 

We can also see that the oscillatory attraction-repulsion interaction relates to the periodic stripy structure of the 
in-plane skyrmion along the $x$-direction, Figure~\ref{fig1}(b) shows how $m_x$ (red) and $m_z$ vary with 
$y$ along lines of $x=0$ (solid curves) and $x=20$nm (dashed curves). One sees clearly an oscillatory behavior 
and the period change with the value of $x$ according to nearly $\lambda_0/\sin\phi$ relation, as shown by 
almost linear line of $1/\lambda$ vs. $\sin\phi$ in the inset with $\phi$ defined above. This proves a nearly 
stripy structure of in-plane skyrmion along the $x$-direction, the same conclusion as those in the main text for 
$H$-induced in-plane skyrmion. 

We can also analytically show that the hidden periodic stripy spin structure is intrinsic for titled symmetric  
magnetic anisotropy and the period is not sensitive to value $K$ as long as the film support isolated titled skyrmions.  
Again, $\Theta$ is defined as the angle between $\mathbf{m}$ and the easy-axis, and $\Phi$ is the angle between 
the $x$-axis and $\mathbf{m}$-component  in the plane perpendicular to the easy-axis. In terms of the nature 
length unit of $4A/\pi D$, $\Theta$ and $\Phi$ of a stable spin texture satisfy the following dimensionless equations,
\begin{equation} 
	\begin{aligned}
		2\nabla^2\Theta-\sin2\Theta\left(\nabla\Phi\right)^2-	\frac{4}{\pi}[\sin\theta\sin2\Theta\partial_y\Phi+2\sin^2\Theta(\cos\Phi\partial_x\Phi+\cos\theta\sin\Phi\partial_y\Phi)]+\frac{2}{\kappa}\sin\Theta\cos\Theta=0
		\label{EquTheta1}
	\end{aligned} 
\end{equation}

\begin{equation} 
	\begin{aligned}
		-\sin2\Theta\left(\nabla\Theta\right)\cdot\left(\nabla\Phi\right)-\sin^2\Theta\nabla^2\Phi+\frac{2}{\pi}[\sin\theta\sin2\Theta\partial_y\Theta
		+2\sin^2\Theta(\cos\Phi\partial_x\Phi+\cos\theta\sin\Phi\partial_y\Phi)]=0,
		\label{EquPhi1}
	\end{aligned}	
\end{equation}

For an in-plane skyrmion ($\theta=\pi/2$) centred at $(0,0)$ with boundary conditions of  
$\Theta(0,0) = 0, ~ \Theta(|x| \to \infty, or |y| \to \infty) = \pi$, Eqs. \eqref{EquTheta1} and \eqref{EquPhi1}
far from the skyrmion center, where $|\pi-\Theta(x,y)| \ll 1$, become
\begin{equation} 
	\begin{aligned}
		\nabla^2\Theta + (\pi-\Theta)\left[\left(\nabla\Phi\right)^2-\frac{4}{\pi}\partial_y\Phi+
		\frac{1}{\kappa}\right] = 0, 
		\label{EquTheta2}
	\end{aligned} 
\end{equation}
\begin{equation} 
	\begin{aligned}
		2(\pi-\Theta)\left[\partial_x\Theta\partial_x\Phi +\partial_y\Theta \left(\partial_y\Phi-\frac{2}{\pi}\right) \right]-(\pi-\Theta)^2(\nabla^2\Phi-2\cos\Phi\partial_x\Theta) = 0. 
		\label{EquPhi2}
	\end{aligned}	
\end{equation}

The same as $H$-controlled anisotropic case, Eqs. \eqref{EquTheta2} and \eqref{EquPhi2} admit following asymptotic solution,
\begin{equation}
	\begin{aligned}
		\Theta(x,y) =\pi- \pi e^{-k_x |x| - k_y |y|},~\Phi(y) = \frac{2}{\pi}y + \Phi_0, 
		\label{Theta}
	\end{aligned}	
\end{equation}
with $k_x^2 + k_y^2 = 1/\kappa - 4/\pi^2$, and $\Phi_0$ is a constant, exactly the same as that in the main text. 

We can also show that skyrmion-skyrmion attraction comes from the match of the periodic stripy structures of 
two in-plane skyrmions in their overlapped region while skyrmion-skyrmion repulsion is due to mismatch of 
structures of them. Same as what was done in the main text, we examine $\Phi$ of two independent skyrmions: 
one centred at $(0,0)$ and  the other at $(x_0, y_0)$. Figure~\ref{fig1}(e) shows $\Phi (l)$ (the red for the 
first skyrmion and the black for the second skyrmion) in the overlapped region for $(x_0, y_0)=(0,136\mathrm{nm} )$, 
$(-50\mathrm{nm},136\mathrm{nm})$, and $(70\mathrm{nm},136\mathrm{nm})$ when two akyrmions attract 
each other, where $l\equiv |\sqrt{x_0^2+y_0^2}(|y|/|y_0|-1/2)$ measures distance from the middle point between 
the two skyrmion along the line of $y=(y_0/x_0)x$.  Indeed, $\Phi (l)$ for two skyrmions are almost identical 
in the overlapped region. They are also the same as the $\Phi (l)$ of the bi-skyrmion molecule. 

In summary, the RKKY-like skyrmion-skyrmion interaction is universal for titled skyrmions and comes from their anisotropic periodic stripe spin structures. These universal features does not depend on the origins of titled skyrmions, no matter whether it is due to a titled magnetic field which is asymmetric or due to the crystalline magnetic anisotropy in which both direction along the easy-axis are equivalent. 

\section*{Boundary equation between the RKKY-like phase and non-isolated skyrmion phase}

In the main text, we have shown that stripy structures of Eq. (6), which gives rise to the RKKY-like interaction, is valid only for $\kappa'>\pi^2/4$ in an in-plane field. To derive the boundary between the RKKY-like phase and non-existence of isolated skyrmion phase in $\kappa'-\theta_H$ parameter space, we go back to Eqs. (4) and (5) in the main text which is reproduced  below,
\begin{equation} 
	\begin{aligned}
		\nabla^2\Theta + (\pi-\Theta)\left[\left(\nabla\Phi\right)^2-\frac{4}{\pi}\sin\theta_H\partial_y\Phi+
		\frac{1}{\kappa'}\right] = 0, 
		\label{EquTheta3}
	\end{aligned} 
\end{equation}
\begin{equation} 
	\begin{aligned}
		2(\pi-\Theta)\left[\partial_x\Theta\partial_x\Phi +\partial_y\Theta \left(\partial_y\Phi-\frac{2}{\pi}\sin\theta_H\right) \right]-(\pi-\Theta)^2(\nabla^2\Phi-2\cos\Phi\partial_x\Phi-2\cos\theta_H\sin\Phi\partial_y\Phi) = 0. 
		\label{EquPhi3}
	\end{aligned}	
\end{equation}
Periodic stripy tail solution of an in-plane skyrmions is 
\begin{equation}
	\begin{aligned}
		\Theta(x,y) =\pi- \pi e^{-k_x |x| - k_y |y|},~\Phi(y) = \frac{2}{\pi}\sin\theta_Hy + \Phi_0, 
		\label{Theta2}
	\end{aligned}	
\end{equation}
where $k_x^2 + k_y^2 = 1/\kappa' - 4\sin^2\theta_H/\pi^2$, and $\Phi_0$ is a constant. 
When $\kappa' <\pi^2/(4\sin^2\theta_H) $, $k_x^2 + k_y^2>0$ the skyrmion is metastable.
When the inequality relation does not hold i.e. $\kappa' >\pi^2/(4\sin^2\theta_H) $, isolated skyrmions does 
not exist which was confirmed by Mumax3 simulations. However,  a conical-like state with $\Theta = const$ 
exists as shown below. Equations (2) and (3) in the main text for $\Theta = const$ become, 
\begin{equation} 
	\begin{aligned}
		\nabla^2\Theta + \sin\Theta\left[\cos\Theta\left(\nabla\Phi\right)^2-\frac{4}{\pi}\sin\theta_H\cos\Theta\partial_y\Phi+
		\frac{1}{\kappa'}\right] = 0, 
		\label{EquTheta4}
	\end{aligned} 
\end{equation}
\begin{equation} 
	\begin{aligned}
		\sin^2\Theta\nabla^2\Phi = 0. 
		\label{EquPhi4}
	\end{aligned}	
\end{equation}
The equations admit following conical solution, 
\begin{equation}
	\begin{aligned}
		\Theta =\arccos\left(\frac{\pi^2}{4\kappa'\sin^2\theta_H}\right),~\Phi(y) = \frac{2}{\pi}\sin\theta_Hy + \Phi_0, 
		\label{Theta3}
	\end{aligned}	
\end{equation}
The energy density of this state is: $-4\sin^2\theta_H[1-\pi^2/(4\kappa'\sin^2\theta_H)]^2/\pi^2<0$.
Thus, the necessary condition for a stable isolated skyrmion is $\kappa'<\pi^2/4\sin^2\theta_H>4$ 
which should also satisfy $\kappa'<4$ \cite{paper7}. In summary, $\kappa'=\pi^2/(4\sin^2\theta_H)$ is the boundary between RKKY-like phase and non-existence isolated skyrmion phase. As shown in the main text, this analytic boundary agrees with Mumax3 simulations well.

\section*{Dependence of RKKY-like features on the type of DMI interaction}

There are two types of DMI interaction: bulk DMI and the interfacial one. Naturally, one may question whether the oscillatory skyrmion interaction and the periodic stripy skyrmion structure still exist if we replace the bulk DMI by an interfacial DMI. Here we show that the general features do not depend on the type of DMI. However, the attractive direction is now perpendicular to skyrmion centre spin direction, instead of parallel. Simultaneously, the stripy tail structures of an isolated skyrmion also rotate a $90\circ$ degree with respected to the bulk DMI case. 
The occurrences of the simultaneous changes in the anisotropic skyrmion interaction and stripy skyrmion tail 
structures further support our assertion of the origin of the RKKY-like interaction. 

\begin{figure}[htbp]
	\centering
	\includegraphics[width=0.96\textwidth]{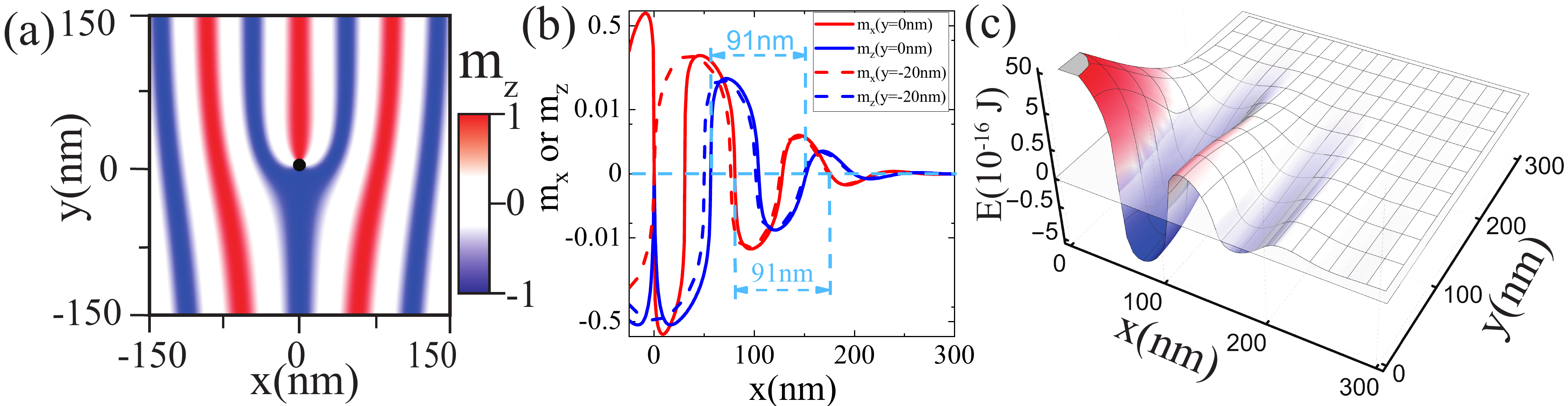}\\
	\caption{(a) Density plots of an in-plane skyrmion with interfacial DMI: $m_z>0$ (red) and $m_z<0$ (blue). 
	The skyrmion center is denoted by the small circle. 
		(b) $m_x$ (red) and $m_z$ (blue) as a function of $x$ for $y=0\,$nm (solid curve) and $20\,$nm 
		(dashed lines). (c) $E(x,y)$. }
	\label{fig2}
\end{figure}

Below, we repeat the simulations presented in the main text for field-induced anisotropy by replacing the bulk DMI with interfacial DMI, $E_{DMI}=\iint m_z\left(\nabla\cdot\mathbf{m}\right)-\left(\mathbf{m}\cdot\nabla\right)m_z\mathrm{d}^2\mathbf{x}$ with the same DMI constant of $D=2\,\mathrm{mJ/m^3}$. All other parameters remain the same as those for Figs. 1-3 in the main text. For $\mathbf{H}=0.164$T parallel to the $-y$-direction and skyrmion centre spin along the $y$-direction, Figure \ref{fig2}(a) is the density plot of $m_z>0$ (red) and $m_z<0$ (blue) to reveal stripy structures along the $x$-direction, perpendicular to the skyrmion centre spin direction. The small circle denotes the skyrmion center. Figure \ref{fig2}(b) plots distributions of $m_x$ and $m_z$ along various direction. The oscillatory distributions reveal the periodic stripy structure of the skyrmion. Figure \ref{fig2}(c) shows 
oscillatory $E(x,y)$ with minima on the $x$-axis, instead of $y$-axis, at $x_{min} = 66.5,\,156.5,\,247.5\,$nm. 
The separation between two neighbouring minimum points is $\lambda\approx156.5-66.5\approx 247.5-156.5\approx 90\sim91\,$nm, and not sensitive to $H$. Interestingly, this period is the same as that in the main text, implying bulk and interfacial DMIs with the same $D$-values are the same on the RKKY-like interaction, except the direction of attractive and repulsive forces.
Thus, two such isolated skyrmions shall attract each other along the $x$-axis. $E(x,y)$ decreases monotonically with $y$ (red line), indicating a repulsive force  between two skyrmions along the $y$-axis. In between, the interaction along other radium, $E(x,y)$, thus interaction, oscillates with a well defined period as explained in the main text.  As shown in the inset of  Fig.~\ref{fig2}(a), the periodic stripes are parallel to centre spin direction (the $y$-direction). Thus, two titled skyrmions attract each other when they align along the $x$-axis (perpendicular to the centre spin direction), and repel each other when they align along $y$-axis. 

The change of attractive direction with the change of type of DMI interaction can be understand from the skyrmion equation, for 
interfacial DMI and $\mathbf{H}$ along the $-y$-direction, 
\begin{equation} 
	\begin{aligned}
		2\nabla^2\Theta-\sin2\Theta_H\left(\nabla\Phi\right)^2-	\frac{4}{\pi}[\sin\theta\sin2\Theta\partial_x\Phi+2\sin^2\Theta(\sin\Phi\partial_x\Phi-\cos\theta_H\cos\Phi\partial_y\Phi)]+\frac{2}{\kappa}\sin\Theta\cos\Theta=0
		\label{EquTheta5}
	\end{aligned} 
\end{equation}

\begin{equation} 
	\begin{aligned}
		-\sin2\Theta\left(\nabla\Theta\right)\cdot\left(\nabla\Phi\right)-\sin^2\Theta\nabla^2\Phi+\frac{2}{\pi}[\sin\theta_H\sin2\Theta\partial_x\Theta
		+2\sin^2\Theta(\sin\Phi\partial_x\Phi-\cos\theta_H\cos\Phi\partial_y\Phi)]=0.
		\label{EquPhi5}
	\end{aligned}	
\end{equation}

Consider an in-plane skyrmion centred at $(0,0)$ and $\Theta(0,0) = 0, ~ \Theta(|x| \to \infty, or |y| \to \infty) = \pi$
 and far from the skyrmion center, $|\pi-\Theta(x,y)| \ll 1$, Eqs. \eqref{EquTheta5} and \eqref{EquPhi5} become
\begin{equation} 
	\begin{aligned}
		\nabla^2\Theta + (\pi-\Theta)\left[\left(\nabla\Phi\right)^2-\frac{4}{\pi}\partial_x\Phi+
		\frac{1}{\kappa}\right] = 0, 
		\label{EquTheta6}
	\end{aligned} 
\end{equation}
\begin{equation} 
	\begin{aligned}
		2(\pi-\Theta)\left[\partial_x\Theta \left(\partial_x\Phi-\frac{2}{\pi}\right)+\partial_y\Theta\partial_y\Phi  \right]-(\pi-\Theta)^2(\nabla^2\Phi-2\cos\Phi\partial_x\Phi) = 0. 
		\label{EquPhi6}
	\end{aligned}	
\end{equation}
The asymptotic solution is 
\begin{equation}
	\begin{aligned}
		\Theta(x,y) =\pi- \pi e^{-k_x |x| - k_y |y|},~\Phi(x) = \frac{2}{\pi}x + \Phi_0, 
		\label{Theta4}
	\end{aligned}	
\end{equation}
where $k_x^2 + k_y^2 = 1/\kappa' - 4/\pi^2$, and $\Phi_0$ is a constant. 
Different from the bulk DMI where $\Phi$ is linear in $y$, $\Phi$ is linear in $x$, 
showing the stripes parallel to the $y$ and leading to an attraction along the $x$-axis.
Similar to the bulk DMI case, $k_x^{-1}$ and $k_y^{-1}$ depends on $\kappa$, 
period $\lambda$ of the periodic stripy structure depends only on the length scale 
$4A/\pi D$ which is  
$\lambda=2\pi \left(\frac{2}{\pi}\right) ^{-1}\times\frac{4A}{\pi D}\approx 93\,$nm.